\documentclass[12pt]{iopart}

\usepackage{latexsym}
\usepackage{graphics}

\usepackage{iopams}
\begin{document}

\title[On strig density at the origin]
{On string density at the origin}

\author{I.S. Kac$^1$ and  V.N. Pivovarchik$^2$}

\address{$^1$  Odessa National Academy of Food Technologies, 112, Kanatnaya str., Odessa, 65039, Ukraine  \\
$^2$ South Ukrainian National Pedagogical University,
Staroportofrankovskaya str., 26, Odessa 65020, Ukraine}
\ead{$^1$ israel@ua.fm, $^2$ v.pivovarchik@paco.net}

\begin{abstract}
In
 [V. Barcilon Explicit solution of the inverse problem for a vibrating string. J. Math. Anal. Appl. {\bf 93} (1983)  222-234] two boundary value problems were considered generated by
the differential equation of a string
$$
y^{\prime\prime}+\lambda p(x)y=0, \ \ 0\leq x \leq L<+\infty 
\eqno{(*)}
$$
with continuous real  function $p(x)$ (density of the string) and the boundary conditions $y(0)=y(L)=0$ the first problem and $y^{\prime}(0)=y(L)=0$ the second one.
In the above paper the following formula was stated 
$$
p(0)=\frac{1}{L^2\mu_1}\mathop{\prod}\limits_{n=1}^{\infty}\frac{\lambda_n^2}{\mu_n \mu_{n+1}}
\eqno{(**)}
$$
where $\{\lambda_k\}_{k=1}^{\infty}$ is the spectrum of the first boundary value problem and $\{\mu_k\}_{k=1}^{\infty}$ of the second one. Rigorous proof of (**) was given in [C.-L. Shen On the Barcilon formula for the string equation with a piecewise continuous density function. Inverse Problems {\bf 21}, (2005) 635--655] under more restrictive conditions of piecewise continuity of $p^{\prime}(x)$. In this paper (**) was deduced using
$$
p(0)=\lim\limits_{\lambda\to +\infty}\left(\frac{\phi(L,-\lambda)}{\lambda^{\frac{1}{2}}\psi(L,-\lambda)}\right)^2
\eqno{(***})
$$
where $\phi(x,\lambda)$ is the solution of (*) which satisfies the boundary conditions $\phi(0)-1=\phi^{\prime}(0)=0$ and
$\psi(x,\lambda)$ is the solution of (*) which satisfies  $\psi(0)=\psi^{\prime}(0)-1=0$.

In our paper we prove that (***) is true for the so-called M.G. Krein's string  which may have any nondecreasing  mass distribution function $M(x)$ with finite nonzero $M^{\prime}(0)$. Also we show that (**) is true for a wide class of strings including those for which $M(x)$ is a singular function, i.e. $M^{\prime}(x)=p(x)\mathop{=}\limits^{a.e.}0$.

\end{abstract}

\vspace{2pc} \noindent{\it Keywords}: spectral function,  Dirichlet boundary condition, Neumann boundary condition, regular string,
singular string, Krein's string.

\submitto{Inverse Problems}
\maketitle

\section{Introduction}
This paper is stimulated by series of papers \cite{B}, \cite{Sh}, \cite{Sh1}  on the so-called Barcilon formula. 
In \cite{B}  V. Barcilon considered the differential equation 
\begin{equation}
\label{1.1}
y^{\prime\prime}+\lambda p(x)y=0,  \ \ 0\leq x\leq L<+\infty,
\end{equation}
where $\lambda$ is the spectral parameter, the function $p(x)$ is continuous and $p(x)>0$ on $[0,L]$. This equation describes small transversal vibrations of a string of linear density $p(x)$ stretched by a unite force, $L$ is the length of the string. In \cite{B} the spectrum of the boundary value problem generated by (\ref{1.1})  and the boundary conditions
$$
y^{\prime}(0)=y(L)=0
$$
was denoted by $\{\mu_n\}_{n=1}^{\infty}$ and the spectrum of the boundary value problem generated by (\ref{1.1})  and the boundary conditions
$$
y(0)=y(L)=0
$$
was denoted by $\{\lambda_n\}_{n=1}^{\infty}$.  There a method of recovering $p(x)$ using $L$,  $\{\mu_n\}_{n=1}^{\infty}$ and $\{\lambda_n\}_{n=1}^{\infty}$ was proposed under a strange condition that the given data are '... such that the solution $p(x)$ of the inverse problem for the vibrating string  exists and is continuous and bounded away from the zero...' (see Theorem 2 in \cite{B}).
In particular, there was established a formula (following C.-L. Shen \cite{Sh} we call it Barcilon formula) expressing $p(0)$ via  $L$,  $\{\mu_n\}_{n=1}^{\infty}$ and $\{\lambda_n\}_{n=1}^{\infty}$:
\begin{equation}
\label{1.4}
p(0)=\frac{1}{L^2\mu_1}\mathop{\prod}\limits_{n=1}^{\infty}\frac{\lambda_n^2}{\mu_n \mu_{n+1}}. 
\end{equation}
The proof in \cite{B} is not correct.
Rigorous proof of this formula  was given by C.-L. Shen \cite{Sh}  under more restrictive conditions of piecewise differentiability of $p(x)$ with $p^{\prime}(x)$ having finite number of discontinuities.

In the context of the proof of (\ref{1.4}) in \cite{Sh}  a formula was obtained equivalent to the following one
$$
p(0)=\lim\limits_{z\to +\infty}\left(\frac{\phi(L,-z)}{z^{\frac{1}{2}}\psi(L,-z)}\right)^2,
$$
where $\phi(x,\lambda)$ and $\psi(x,\lambda)$ ($0\leq x\leq L$)  denote solutions of a differential equation more general than (\ref{1.1}) (see (\ref{3.2}) below) which satisfy the  conditions $\phi(0,\lambda)=\psi_+^{\prime}(0,\lambda)=1$, $\phi_+^{\prime}(0,\lambda)=\psi(0,\lambda)=0$.

The aim of this paper is to express the value at $x=0$ of the right derivative of the mass distribution function $M(x)$ of the string via 
the same data  $L$,  $\{\mu_n\}_{n=1}^{\infty}$ and $\{\lambda_n\}_{n=1}^{\infty}$ not imposing any restrictions of continuity on
the density $p(x)=M^{\prime}(x)$. Moreover, for a wide class of strings we prove Barcilon formula (\ref{1.4}) despite $\frac{dM}{dx}\mathop{=}\limits^{a.e.}0$ on $[0,L]$ (see Theorem 5.2 and 5.3). Since $M(x)\not=const$, this  means that $M(x)$ is not continuous.  In this purpose we need first to describe some results on the spectral theory of the string with a regular left end (M. G. Krein's string) and, for this, some results on the so-called $R$-functions.

\section{Classes of functions ($R$),  ($S$) and ($S^{-1}$)}

\setcounter{equation}{0}  

\subsection{Class ($R$)}

According to the terminology of \cite{KK1} a function $f(z)$ of a complex variable $z$ is said to be an {\it $R$-function} or to belong to  the class (R) if 

1) it is defined and holomorphic in each of the half-planes ${\rm Im} z>0$ and  ${\rm Im} z<0$,

2) $f(\overline{z})=\overline{f(z)}$ (${\rm Im} z\not=0$),

3) ${\rm Im}z \ {\rm Im} f(z)\geq 0$ (${\rm Im} z\not=0$). 

Such a function is also often called {\it Nevanlinna} function.

It is easy to see that an $R$-function $f(z)$ can attain a real value for nonreal $z$  if and only if it is a real constant.  Such an $R$-function is said to be {\it degenerate}. A nondegenerate $R$-function maps the open upper (lower) half-plane into itself. 

A function $f$ is an $R$-function if and only if it can be represented as (see \cite{KK1}) 
\begin{equation}
\label{2.1}
f(z)=\alpha+\beta z+\int_{-\infty}^{\infty}\left(\frac{1}{\lambda-z}-\frac{\lambda}{1+\lambda^2}\right)d\tau(\lambda),
\end{equation}  
where $\beta\geq 0$, $\alpha\in \mathbb{R}$ and $\tau$ is a nondecreasing function such that
$$
\int_{-\infty}^{\infty}\frac{d\tau(\lambda)}{1+\lambda^2}<\infty,
$$
what guarantees absolute convergence of the integral in (\ref{2.1}).
We will  normalize $\tau$ in representation (\ref{2.1}) of an $R$-function (or a function belonging to a subclass of ($R$), see below) as follows
$$
\tau(\lambda)=\frac{1}{2}(\tau(\lambda+0)+\tau(\lambda-0)) \ \ \forall \lambda\in\mathbb{R}, \ \ \tau(0)=0.
$$
Under such normalization $\tau(\lambda)$ which we will call the {\it spectral function for $R$-function $f(z)$} is uniquely determined by
the Stietjes inversion formula
$$
\tau(\lambda)=\frac{1}{\pi}\lim\limits_{\epsilon\to+0}\int_0^{\lambda}{\rm Im} f(t+i\epsilon)dt  \ \  \forall \ \ \lambda\in\mathbb{R}.
$$
The constant $\beta$ can be obtained as
$$
\beta=\lim\limits_{\eta\to+\infty}\frac{{\rm Im} f(i\eta)}{\eta}.
$$
Each $R$-function consists of two functions one of which (upper part) is holomorphic in the open upper half-plane while the other (lower part) is holomorphic in the lower half-plane. In general case the lower part is not an analytic continuation of the upper part. However, if there exists an interval $(a,b)$ where the spectral function $\tau(\lambda)$ is constant then the integral in the right-hand side of (\ref{2.1}) exists not only for nonreal $z$ but also for $z\in (a,b)$ and    attains real values there. In this case the lower part of the $R$-function is the analytic continuation of the upper part and this continuation attains  real values on $(a,b)$. In the sequel we will deal only with $R$-functions defined not only for $\mathbb{C}\backslash\mathbb{R}$ but also on  intervals of $\mathbb{R}$ where $\tau(\lambda)$ is constant. It should be mentioned that each nondegenerate $R$-function is monotonically increasing on the intervals where its spectral function is constant.  Also it is known (see \cite{KK1}, Sec. 2, Subsec. 3)
 that if in all points $\lambda\in (a,b)\subset\mathbb{R}$ 
$$
\lim\limits_{\epsilon\to +0}\frac{1}{\pi}{\rm Im} f(\lambda+i\epsilon)=g(\lambda)
$$
and the function $g(\lambda)$ is bounded on $(a,b)$ then the spectral function $\tau(\lambda)$ of the $R$-function $f(\lambda)$ is absolutely continuous on $(a,b)$ and $\tau^{\prime}(\lambda)\mathop{=}\limits^{a.e.}g(\lambda)$. 

\subsection{Class ($S$)}

A function $f$ is said to be an {\it $S$-function} or to belong to the class ($S$) if 

1) $f\in (R)$,

2) $f$ is holomorphic in ${\rm Ext}[0,\infty)$  ($=\mathbb{C}\backslash[0,\infty)$),

3) $f(z)\geq 0$ for all $z\in (-\infty,0)$.

As it follows from Subsection 2.1 the spectral function $\tau(\lambda)$ of an $S$-function $f(\lambda)$ is constant on $(-\infty,0)$. Moreover,
it is known (\cite{KK1}, Sec.5, Theorem S.1.5.1) that
$$
\int_{-0}^{\infty}\frac{d\tau(\lambda)}{1+\lambda}<\infty,
$$
and  representation (\ref{2.1}) can be reduced to
\begin{equation}
\label{2.7}
f(z)=\gamma+\int_{-0}^{\infty}\frac{d\tau(\lambda)}{\lambda-z}
\end{equation}
where $\gamma\geq 0$.  

\subsection{Class ($S^{-1}$)}

A function $f$ is said to be an {\it $S^{-1}$-function} or to belong to the class ($S^{-1}$) if 

1) $f\in (R)$,

2) $f$ is holomorphic in ${\rm Ext}[0,\infty)$  ($=\mathbb{C}\backslash[0,\infty)$),

3) $f(z)< 0$ for all $z\in (-\infty,0)$.

As it follows from Subsection 2.1 the spectral function $\tau(\lambda)$ of an $S^{-1}$-function $f(z)$ is constant on $(-\infty,0)$ and Theorem S.1.5.2 in \cite{KK1} (\cite{KK1}, Sec.5) implies $\tau(+0)=\tau(-0)$ and
$$
\int_{-0}^{\infty}\frac{d\tau(\lambda)}{\lambda+\lambda^2}<\infty.
$$
Representation (\ref{2.1}) can be reduced to 
$$
f(z)=\alpha+\beta z+\int_{+0}^{\infty}\left(\frac{1}{\lambda-z}-\frac{1}{\lambda}\right) d\tau(\lambda)
$$
where $\alpha\leq 0$.
It was shown in proof of Theorem S.1.5.2 in \cite{KK1} that if $f(z)$ is holomorphic at $z=0$ then $\alpha<0$ if and only if 
$
f(0)\not=0.
$


\section{Main  theses of the spectral theory of strings}

\setcounter{equation}{0}

\subsection{Sting and classification of its ends}

Let $I$ be an interval of one of the kinds $(a,b)$, $(a,b]$, $[a,b)$ or $[a,b]$ where $-\infty\leq a<b\leq\infty$ (in the case of $a\in I$ or $b\in I$ we have $a>-\infty$ and $b<+\infty$, respectively). Let $M(x)$  ($M(x)<\infty$ for all $x\in I$)  be a  nondecreasing function on $I$ which can have jumps, intervals of constant value, absolutely continuous, singular continuous parts. We set $a_0=: inf {\cal F}_M$ ,  $b_0=: sup {\cal F}_M$, where ${\cal F}_M$ is the set of points of growth  of $M(x)$. We assume that ${\cal F}_M$  is an infinite set of points. Let us associate with $I$ and $M$ 
a string $S(I,M)$ with the mass distribution described by $M(x)$ in  sense that $M(x_2+0)-M(x_1-0)$ is the mass of the part of the string located on $[x_1,x_2]$ for each $x_1, x_2\in I$ and $x_1\leq x_2$ ( here we assume $M(a-0)=M(a)$ if $a\in I$ and $M(b+0)=M(b)$ if $b\in I$). The left (right) end of the string $S(I,M)$ is said to be {\it regular} if $a_0>-\infty$, $M(a)>-\infty$ ($b_0<\infty$, $M(b)<\infty$).     
In the opposite case the end is said to be {\it singular}. A regular end is said to be {\it completely regular} if $a\in I$ ($b\in I$). A string with both ends regular is said to be a {\it regular string}. In the opposite case it is said to be a {\it singular string}. If $a\notin
I$ ($b\notin I$) we set
$$
M(a)=\mathop{\lim}\limits_{x\to a+0}M(x) \ \  (M(b)=\mathop{\lim}\limits_{x\to b-0}M(x)).
$$

\subsection {Differential operation and differential equation of a sting}

In this paper we are dealing with  strings  each  completely regular end of which does not bear a point mass. Therefore,
we use a definition of the differential operation $l_{I,M}$ of the string $S(I,M)$ which fits to this situation.

{\bf Remark} {\it In the general case one should introduce a notion of associate derivatives: the 'left derivative' $f_-^{\prime}(a)$ if the left end of $I$ is completely regular and the 'right derivative' $f_+^{\prime}(b)$ if the right end is completely regular. }

{\bf Definition 3.1} {\it  Let ${\cal D}_{M,I}$ be the set of all functions $f(x)$ defined on $I$ such that \\
1) $f$ is locally absolutely continuous on $I$, 2) there exist finite left $f^{\prime}_-(x)$ and right $f^{\prime}_+(x)$ derivatives at each interior point $x$ of $I$, 3) there exists $M$-measurable function $g$ such that for any two points $x_1, x_2\in I$ ($x_1\leq x_2$)
\begin{equation}
\label{3.1} 
f_{\pm}^{\prime}(x_2)-f_{\pm}^{\prime}(x_1)=-\int_{x_1\pm 0}^{x_2\pm 0}g(x)dM(x)
\end{equation}
for each of four combinations of signs the same in both sides of (\ref{3.1}) for the same $x_j$ ($j=1,2$). 

For a function $f\in {\cal D}_{I,M}$ we set
$l_{M,I}[f](x)=g(x)$ where $g$ is the function involved in (\ref{3.1}).}

{\bf Remark} {\it We  have defined $l_{M,I}[f](x)$ up to equivalence with respect to the $M$- measure. It is clear that for $f\in {\cal D}_{I,M}$ 
$$
l_{I,M}[f](x)=-\frac{d}{(d)M(x)}f_+^{\prime}(x)=-\frac{d}{(d)M(x)}f_-^{\prime}(x)
$$
at $M$- almost all $x$. 
Here $\frac{d}{(d)M(x) }$ is the symbol of the symmetric derivative with respect to $M$.}

We call {\it differential equation of the string $S(I,M)$ }
\begin{equation}
\label{3.2}
l_{I,M}[y]-\lambda y=0 \ \ (x\in I),
\end{equation}
where $\lambda$  is the spectral parameter. A function $u\in {\cal D}_{I,M}$ is s said to be a solution of (\ref{3.2}) if
$
l_{I,M}[u](x)-\lambda u(x)=0
$
for $M$-almost all $x\in I$.   

\subsection{M. G. Krein's strings $S_1(I,M)$ and $S_0(I,M)$}

We deal with strings $S(I,M)$ the left  ends of which are completely regular (for convenience we place them at $x=0$) while the right ends $x=L$ are either regular and then completely regular with $L<+\infty$ or singular and then $L\leq +\infty$ and $L\notin I$. In present 
paper we assume that $inf {\cal F}_M=0$ and, as it was mentioned above, completely regular ends of strings do not bear point masses.
 
A string $S(I,M)$ is said to be $S_1(I,M)$ if its left end is free to move without friction in the direction orthogonal to $x$-axis, i.e. to the equilibrium position of the string. By $S_0(I,M)$ we denote a string $S(I,M)$ with the left end fixed. We assume that if the right end of a string $S_1(I,M)$ or $S_0(I,M)$ is (completely) regular then it is fixed.

We define {\it fundamental} functions $\phi(x,\lambda)$ and $\psi(x,\lambda)$ of strings $S_1(I,M)$ and  $S_0(I,M)$, respectively, as the solutions of equation (\ref{3.2})  which satisfy the initial conditions $\phi(0,\lambda)=1$, $\phi^{\prime}_+(0,\lambda)=0$ and $\psi(0,\lambda)=0$, $\psi^{\prime}_+(0,\lambda)=1$. 

It is  known (see \cite{KK}, Sec. 2 )  that for any fixed $x\in I$ the functions $\phi(x,\lambda)$, $\psi(x,\lambda)$, $\phi^{\prime}_-(x,\lambda)$, $\psi^{\prime}_-(x,\lambda)$,  $\phi_+^{\prime}(x.\lambda)$ and $\psi_+^{\prime}(x,\lambda)$ are entire real functions of $\lambda$ of order not more than $1/2$. 

{\bf Remark} {\it  A meromorphic in $\mathbb{C}$ or an  entire function is said to be real if it  attains real values for real values of variable.}

Since $\phi(x,0)=1$ and $\psi(x,0)=x$, for each fixed $x\in I$ the following representations are valid:
$$
\phi(x,\lambda)=\prod\limits_j\left(1-\frac{\lambda}{\mu_j(x)}\right),  \ \  \psi(x,\lambda)=x\prod\limits_j\left(1-\frac{\lambda}{\lambda_j(x)}\right), 
$$
where $\mu_j(x)$, $\lambda_j(x)$, $j=1,2,...$ are zeros of entire in $\lambda$ functions $\phi(x,\lambda)$ and $\psi(x,\lambda)$, respectively.  

Usually, the set of squares of frequencies of free vibrations of a regular string is called its {\it spectrum}. The spectrum depends on the mass distribution and on the ways of connection of its ends. Therefore, we mean by spectra of strings $S_1(I,M)$ and $S_0(I,M)$ the sets of eigenvalues of the boundary value problems
\begin{equation}
\label{3.4}
l_{I,M}[y]-\lambda y=0, \ \ y^{\prime}_+(0)=y(L)=0,
\end{equation}
\begin{equation}
\label{3.5}
l_{I,M}[y]-\lambda y=0, \ \ y(0)=y(L)=0,
\end{equation}
respectively. It is easy to see that the spectrum $\{\mu_j\}_{j=1}^{\infty}$  of  a completely regular string $S_1(I,M)$ is the set of zeros of the entire function $\phi(L,\lambda)$ and the spectrum $\{\lambda_j\}_{j=1}^{\infty}$ of the completely regular string  $S_0(I,M)$ is the set of zeros of $\psi(L,\lambda)$
with the same $I$ and $M$, respectively. This is in accordance with a general definition of $S_0(I,M)$ and $S_1(I,M)$ strings spectra given below.

\subsection{Spectral functions of strings $S_1(I,M)$  and the coefficient of dynamic compliance}

Let us denote by ${\cal L}_M^2(I)$ the set of $M$-measurable functions $f(x)$ such that
$$
\|f\|_M^2:=\int_I|f(x)|^2dM(x)<\infty.
$$
We denote by  $  \hat{{\cal L}}_M^2(I)$ the set of functions $f\in {\cal L}_M^2(I)$ which are identically zero in some neighborhood of $x=L$ if the right end is singular.  If the right end is regular then $ \hat{ \cal{L}}_M^2(I)= {\cal L}_M^2(I)$.

{\bf Definition 3.1} {\it A nondecreasing on $(-\infty,\infty)$ function $\tau(\lambda)$ normalized by the conditions 
$$
\tau(\lambda)=\frac{1}{2}(\tau(\lambda+0)+\tau(\lambda-0)) \ \forall  \lambda\in(-\infty,\infty), \ \ \tau(0)=0,
$$
is said to be a spectral function of the string $S_1(I,M)$ ($S_0(I,M)$) if the mapping $U: f\to {\cal F}$ where $f\in \hat{{\cal L}}_M^2(I)$ and 
$$
{\cal F}(\lambda)=\int_0^Lf(x)\phi(x,\lambda)dM(x) \ \ ({\cal F}(\lambda)=\int_0^Lf(x)\psi(x,\lambda)dM(x))
$$
maps isometrically $ \hat{{\cal L}}_M^2(I)$ into ${\cal L}_{\tau}^2(-\infty,\infty)$, i.e. if for each function $f\in \hat{\cal L}_M^2(I)$ the 'Parceval identity' is true:
$$
\int_{-\infty}^{\infty}|{\cal F}(\lambda)|^2d\tau(\lambda)=\int_I|f(x)|^2dM(x),
$$
where ${\cal F}=Uf$. A spectral function is said to be orthogonal if  $U$ maps $ \hat{{\cal L}}_M^2(I)$ into a dense part of  ${\cal L}_{\tau}^2(-\infty,\infty)$. The set of points of growth of a spectral function is said to be the spectrum of it.}

The function
\begin{equation}
\label{3.7}
T(z):=\mathop{\lim}\limits_{x\to L-0}\frac{\psi(x,z)}{\phi(x,z)}, \ \  z\in(\mathbb{C}\backslash [0,+\infty) )
\end{equation} 
is said to be the {\it coefficient of dynamic compliance} of the string $S_1(I,M)$. If the right end $x=L$ is completely regular then $T(z)=\frac{\psi(L,z)}{\phi(L,z)}$ is a meromorphic function. In any case $T(z)$ is an  $S$-function.

Being an $S$-function  $T(z)$ has the spectral function $\tau^{(1)}(\lambda)$ which is constant on $(-\infty,0)$.  Since $\inf {\cal F}_M=0$, we have  (see \cite{KK1}, Sec.5, \cite{KK}, Sec.10) instead of general representation (\ref{2.7}): 
\begin{equation}
\label{3.8}
T(z)=\int_{-0}^{+\infty}\frac{d\tau^{(1)}(\lambda)}{\lambda-z}, \ \  z\in(\mathbb{C}\backslash [0,+\infty) ),
\end{equation}
 We keep  the norming 
$$
\tau^{(1)}(\lambda)=\tau^{(1)}(-0) \ \ {\rm for} \ \ \lambda<0, 
$$
\begin{equation}
\label{3.9}
 \tau^{(1)}(\lambda)=\frac{1}{2}(\tau^{(1)}(\lambda+0)+\tau^{(1)}(\lambda-0)) \ \ \forall \lambda\in\mathbb{R}, \ \
\tau^{(1)}(0)=0, 
\end{equation}
usual for $R$-functions (remind that $(S)\subset (R)$).
 Being normed this way the function $\tau^{(1)}(\lambda)$ is a   spectral function of the string  $S_1(I,M)$  
(see \cite{KK}, Sec.3 Main Theorem and Sec.10, Theorem 10.1).  
This spectral function of the string $S_1(I,M)$  is called {\it main spectral function}. The main spectral function is orthogonal. Notice that
$$
\int_{-0}^{+\infty}\frac{d\tau^{(1)}(\lambda)}{1+\lambda}<\infty. 
$$
In case of singular string, $\tau^{(1)}(\lambda)$ is its unique spectral function with nonnegative spectrum. The spectrum of the main spectral function of a string $S_1(I,M)$ is said to be the spectrum of this string.

\subsection{The main spectral function of the string $S_0(I,M)$}

The function $-\frac{1}{T(z)}$ is an $S^{-1}$-function. In our case it has a unique representation of the form (\cite{KK}, Sec. 12)
$$
-\frac{1}{T(z)}=-\frac{1}{L}+\int_{+0}^{+\infty}\left(\frac{1}{\lambda-z}-\frac{1}{\lambda}\right)d\tau^{(0)}(\lambda)
$$  
where $\tau^{(0)}(\lambda)$ is a  nondecreasing function, normalized by (\ref{3.9}) with $\tau^{(0)}(\lambda)$ instead of $\tau^{(1)}(\lambda)$. This function is a spectral function of the string $S_0(I,M)$  generated by the same string $S(I,M)$. It is said to be the {\it main spectral function} of this string $S_0(I,M)$ and its spectrum to be the spectrum of this string.

\subsection{Kasahara's theorem}

It was shown in \cite{Kac} that if a string $S_1(I,M)$ is such that for some fixed $\alpha>0$ there exists a nonzero finite limit
\begin{equation}
\label{3.12}
\lim\limits_{x\to+0}\frac{M(x)}{x^{\alpha}}
\end{equation}
then there exist the limits
\begin{equation}
\label{3.13}
\lim\limits_{z\to+\infty}T(-z)z^{\frac{1}{\alpha+1}}, \ \  \lim\limits_{\lambda\to +\infty}\tau^{(1)}(\lambda)\lambda^{-\frac{\alpha}{\alpha+1}}
\end{equation}
and the following is true
\begin{equation}
\label{3.14}
\lim\limits_{x\to+0}\frac{M(x)}{x^{\alpha}}=\left(B(\alpha)\Gamma\left(\frac{1}{\alpha+1}\right)\Gamma\left(\frac{2\alpha+1}{\alpha+1}\right)\right)^{\alpha+1}\left(\lim\limits_{z\to+\infty}T(-z)z^{\frac{1}{\alpha+1}}\right)^{-(\alpha+1)}.
\end{equation}
$$
=\left(B^{-1}(\alpha)\lim\limits_{\lambda\to+\infty}\tau^{(1)}(\lambda)\lambda^{-\frac{\alpha}{\alpha+1}}\right)^{-(\alpha+1)}
$$
where $\Gamma$ is Euler's gamma-function and
\begin{equation}
\label{3.15}
B(\alpha)=\left(\frac{\alpha}{(\alpha+1)^2}\right)^{\frac{\alpha}{\alpha+1}}\Gamma^{-2}\left(\frac{2\alpha+1}{\alpha+1}\right).
\end{equation}
For $\alpha=1$ equations (\ref{3.14})  can be reduced to
\begin{equation}
\label{3.16}
\lim\limits_{x\to+0}\frac{M(x)}{x}=\left(\lim\limits_{z\to+\infty}(T(-z)z^{\frac{1}{2}})\right)^{-2}=
\left(\frac{\pi}{2}\lim\limits_{\lambda\to+\infty}\tau^{(1)}(\lambda)\lambda^{-\frac{1}{2}}\right)^{-2}.
\end{equation}
{\bf Remark} {\it Equations  (\ref{3.14})  remain true if $\tau^{(1)}(\lambda)$ is changed for any other spectral function of the same string (see \cite{Kac}, Lemma 6). }
  
In the case of a string $S(I,M)$ with a regular right end we have 
$T(z)=\frac{\psi(L,z)}{\phi(L,z)}$.
 Therefore, the first equation in (\ref{3.16}) is an analogue of the one  which is also called {\it Barcilon} in \cite{Sh1}.

Some years after \cite{Kac}, a paper  by Kasahara \cite{Kas} appeared where the results of \cite{Kac} were generalized and inverted. In particular, Theorem 2 in \cite{Kas} implies that existence and being finite and nonzero of any of the  limits in (\ref{3.13}) guarantees existence of another limit in (\ref{3.13}) and of the limit in (\ref{3.12}) and also validity of  (\ref{3.14}). For $\alpha=1$ it means that 
if one of the three limits in (\ref{3.16}) exists and is finite and nonzero then the two other also exist and (\ref{3.16}) is true.

It should be noticed that in contradistinction to  \cite{B}, \cite{Sh}, \cite{Sh1} the results in \cite{Kac} and \cite{Kas} were obtained without any assumption of continuity or piecewise differentiability of the  density of the string.  By the way, the first limit in (\ref{3.16}), i.e. the right derivative of the function $M(x)$ at $x=0$, i.e. the density of the string at $x=0$ can exist and be finite and nonzero even in the case where $M(x)$ is a pure jump function. Example of such a function can be easily constructed.

\subsection{Main spectral function of the string $S_1(I,M)$  and its length (length of the  \ \  interval $I$)}

If $\tau^{(1)}(\lambda)$ is the main spectral function of a string $S_1(I,M)$ (regular or singular)  then 
\begin{equation}
\label{3.18}
\int_{-0}^{+\infty}\frac{d\tau^{(1)}(\lambda)}{\lambda}=L
\end{equation} 
in both cases of finite and infinite $L$.
This result was obtained by M.G. Krein \cite{Kr}.

 It should be mentioned that if for some $\epsilon>0$ the interval $[0,\epsilon)$ has zero $\tau^{(1)}$-measure then  the integral in (\ref{3.18}) is finite and (\ref{3.8}) implies $T(0)=L<\infty$.

\section{Relation between the discrete spectra of the strings $S_1(I,M)$ and $S_0(I,M)$  generated by the same string $S(I,M)$ and behavior of $M(x)$ at $x\to +0$ }

\subsection{Spectral function via two spectra}

\setcounter{equation}{0}

Let a string $S(I,M)$ have the length $L<\infty$, the string $S_1(I,M)$ generated by  $S(I,M)$ have discrete spectrum $\{\mu_k\}_{k=1}^{\infty}$ where $0<\mu_1<\mu_2<...$ and let $\{\lambda_k\}_{k=1}^{\infty}$ where $\lambda_1<\lambda_2<...$ be the spectrum of the string $S_0(I,M)$ generated by the same string $S(I,M)$. It is known that these spectra interlace:
$$
0<\mu_1<\lambda_1<\mu_2<\lambda_2<...
$$    
Kasahara's theorem mentioned above shows that information about the behavior of $M(x)$ at $x\to+0$ can be extracted from behavior of the main spectral function  $\tau^{(1)}(\lambda)$ of the string $S_1(I,M)$ at $\lambda\to +\infty$. 
To express $\tau^{(1)}(\lambda)$  through $L$, $\{\mu_k\}_{k=1}^{\infty}$ and $\{\lambda_k\}_{k=1}^{\infty}$  we will use the following theorem:

{\bf Theorem 4.1}
{\it The function $T(z)$ defined by (\ref{3.7}) admits representation
\begin{equation}
\label{4.1}
T(z)=L\prod\limits_{k=1}^{\infty}\frac{1-\frac{z}{\lambda_k}}{1-\frac{z}{\mu_k}}, \ \  z\in(\mathbb{C}\backslash\{\mu_k\}_{k=1}^{\infty})
\end{equation}
independently of whether the string is regular or not.}

{\bf Proof} First of all we notice that  (\ref{3.8}) attains the form
\begin{equation}
\label{4.2}
T(z)=\sum\limits_{k=1}^{\infty}\frac{\rho_k}{\mu_k-z}
\end{equation}
where $\rho_j>0$ are the jumps of $\tau^{(1)}(\lambda)$ at $\mu_j$ ($j=1,2,...$).
This implies that the coefficient of dynamic compliance $T(z)$ of the string $S_1(I,M)$ which in general is defined on $\mathbb{C}\backslash [0,+\infty)$, in our case can be extended by continuity onto the intervals $(\mu_j,\mu_{j+1})$ and attains real values on these intervals. After this $T(z)$ appears to be holomorphic in the domain $\mathbb{C}\backslash\{\mu_k\}_{k=1}^{\infty}$ (see \cite{KK1} or Subsection 2.1). 

It follows from Subsections 3.4 and 3.5 that the set of points of the spectrum of the string $S_0(I,M)$ coincides with the set of zeros of the function $T(z)$. We notice that according to (\ref{3.8}), (\ref{3.18}) and (\ref{4.2})
$$
L=\int_{-0}^{+\infty}\frac{d\tau^{(1)}(\lambda)}{\lambda}=\sum\limits_{j=1}^{\infty}\frac{\rho_j}{\mu_j}=T(0),
$$
and $0<T(z)<L \ \forall \ z\in (-\infty,0)$. Moreover, $T(z)$ is continuous and monotonically increasing on $(-\infty,\mu_1]$. On each of the intervals $(\mu_j, \mu_{j+1})$ the function $T(z)$ is continuous and monotonically increases from $-\infty$ to $+\infty$. It is clear that $T(z)<0$ for $z\in (\mu_j,\lambda_j)$ and  $T(z)>0$ for $z\in (\lambda_j, \mu_{j+1})$, $j=1,2,...$. 
 Let us set 
\begin{equation}
\label{4.4}
Q(z):=\frac{1}{L}T(z).
\end{equation}
It is clear that $Q(z)$ is also an $S$-function  which is holomorphic in $\mathbb{C}\backslash \{\mu_j\}_{j=1}^{\infty}$, attains real values of the same sign as $T(z)$ on $(-\infty,\mu_1)$, $(\mu_j, \mu_{j+1})$ ($j=1,2,...$). Also it is clear that $0<Q(z)<1$ for $z\in(-\infty,0)$ and $Q(0)=1$. 

Let us consider the function
\begin{equation}
\label{4.5}
U(z)=\log Q(z),
\end{equation}
where by $\log$ we mean  the branch of $w=\log v$  defined in $\mathbb{C}\backslash (-\infty,0]$ which attains real values for $v\in (0,+\infty)$. Since $\log v=\log|v|+i \arg v$ this means that $\arg v=0$ for $v\in (0,+\infty)$ and due to continuity of $\log v$ we have $0<\arg v<\pi$ for ${\rm Im}v>0$ and $-\pi<\arg v<0$ for ${\rm Im} v<0$. Therefore, for the considered branch of $\log v$ 
the inequality ${\rm Im} v  \ {\rm Im}\log v>0$ is true for ${\rm Im} v\not=0$ and $\log\overline{v}=\overline{\log v}$. Thus, our branch of $\log v$ is an $R$-function.

 Since $0<Q(z)<Q(0)=1$  for $z<0$, this implies $U(z)<0$ for $z<0$. Thus, $U(z)$ is an $S^{-1}$-function and admits the representation 
\begin{equation}
\label{4.6}
U(z)=\alpha+\beta z+\int_{+0}^{+\infty}\left(\frac{1}{\lambda-z}-\frac{1}{\lambda}\right)d\sigma(\lambda)
\end{equation}
 where $\alpha\leq 0$, $\beta\geq 0$ and $\sigma(\lambda)$ being the spectral function of an $R$-function  is nondecreasing on
$(0,+\infty)$ and $\beta=\lim\limits_{\eta\to +\infty}\frac{{\rm Im}U(i\eta)}{\eta}$. In our case $0<{\rm Im} U(i\eta)= \arg Q(i\eta)<\pi$ for all $\eta>0$ and, therefore, $\beta=0$.
 
As it was mentioned in Subsection 2.3 $\alpha\not=0$ if and only if $U(0)\not=0$. However, in our case $Q(0)=1$ and this implies $U(0)=0$ and therefore $\alpha=0$.  Now it remains to clarify what $\sigma(\lambda)$ in (\ref{4.6}) is in our case. 

As it was mentioned in Subsection 2.1 if an $R$-function (and consequently an $S^{-1}$-function) $f(z)$ is such that for each $\lambda\in(a,b)$: $\frac{1}{\pi}\lim\limits_{\epsilon\to+0}f(\lambda+i\epsilon)=g(\lambda)$ and $g(\lambda)$ is bounded on $(a,b)$ then the spectral function of  $f(z)$ is absolutely continuous and its derivative equals $g(\lambda)$ almost everywhere on $(a,b)$.

If $\lambda\in(\lambda_j,\mu_{j+1})$ then $Q(\lambda)>0$ and $\arg Q(\lambda)=0$. Thus, ${\rm Im} U(\lambda)=0 \ \forall \ \lambda\in (\lambda_j, \mu_{j+1})$ and it follows from the above mentioned  that $\sigma(\lambda)$   is absolutely continuous on $(\lambda_j,\mu_{j+1})$ ($j=1,2,...$) and $\sigma^{\prime}(\lambda)\mathop{=}\limits^{a.e.}0$  and, consequently, $\sigma^{\prime}(\lambda)=0$ for all $\lambda\in (\lambda_j,\mu_{j+1})$.

For $\lambda\in (\mu_j,\lambda_j)$ we  have $Q(\lambda)<0$ and, consequently, $\lim\limits_{\epsilon\to +0}\arg Q(\lambda+i\epsilon)=\pi  \ \forall \lambda\in (\mu_j,\lambda_j$). Thus,  $\lim\limits_{\epsilon\to+0}{\rm Im} U(\lambda+i\epsilon)=\pi \ \forall \lambda\in (\mu_j,\lambda_j)$. Therefore, $\sigma(\lambda)$ is absolutely continuous on $(\mu_j,\lambda_j)$ and $\sigma^{\prime}(\lambda)\mathop{=}\limits^{a.e.}1$ on $(\mu_j,\lambda_j)$. This implies $\sigma^{\prime}(\lambda)=1 \ \forall \lambda\in (\mu_j,\lambda_j)$, $(j=1,2,...)$.

Thus, equation (\ref{4.6}) attains the form
\begin{equation}
\label{4.7}
U(z)=\sum\limits_{j=1}^{\infty}\int_{\mu_j}^{\lambda_j}\left(\frac{1}{\lambda-z}-\frac{1}{\lambda}\right)d\lambda.
\end{equation} 
The series in (\ref{4.7}) converge
for all nonreal $z$ because the integral in (\ref{4.6}) converges. 

It is clear that
$$
\int_{\mu_j}^{\lambda_j}\left(\frac{1}{\lambda-z}-\frac{1}{\lambda}\right)d\lambda=\log\frac{1-\frac{z}{\lambda_j}}{1-\frac{z}{\mu_j}}.
$$
Substituting it into (\ref{4.7}) we obtain
$$
U(z)=\sum\limits_{j=1}^{\infty}\log\frac{1-\frac{z}{\lambda_j}}{1-\frac{z}{\mu_j}}
$$
and due to (\ref{4.4}) and (\ref{4.5})
$$
T(z)=LQ(z)=Le^{U(z)}=L \ exp \sum\limits_{j=1}^{\infty}\log\frac{1-\frac{z}{\lambda_j}}{1-\frac{z}{\mu_j}}=
$$
$$
L  \prod\limits_{j=1}^{\infty} exp \left(\log\frac{1-\frac{z}{\lambda_j}}{1-\frac{z}{\mu_j}}\right)=L\prod\limits_{j=1}^{\infty}\frac{1-\frac{z}{\lambda_j}}{1-\frac{z}{\mu_j}}.
$$
Theorem is proved.

Equation (\ref{4.1})
implies $\rho_j=- \mathop{res}_{\mu_j}T(z)$ and using (\ref{4.2}) we obtain
$$
\rho_j=L(\lambda_j-\mu_j)\frac{\mu_j}{\lambda_j}\prod\limits_{k\not=j}\frac{1-\frac{\mu_j}{\lambda_k}}{1-\frac{\mu_j}{\mu_k}}.
$$

Thus, 
\begin{equation}
\label{4.9}
\tau^{(1)}(\lambda+0)=\sum\limits_{\mu_j\leq\lambda}\rho_j=
L\mathop{\sum}\limits_{\mu_j\leq\lambda}(\lambda_j-\mu_j)\frac{\mu_j}{\lambda_j}\prod\limits_{k\not=j}\frac{1-\frac{\mu_j}{\lambda_k}}{1-\frac{\mu_j}{\mu_k}}.
\end{equation}

{\bf Theorem 4.2}  {\it Let two sequences  $\{\mu_n\}_{n=1}^{\infty}$ and $\{\lambda_n\}_{n=1}^{\infty}$ 
 interlace:
\begin{equation}
\label{4.10}
0<\mu_1<\lambda_1<\mu_2<\lambda_2<...
\end{equation}
 
Then there exists a unique string $S(I,M)$ of a given finite length $L$ such that the spectrum of the string $S_1(I,M)$ generated by $S(I,M)$   coincides with $\{\mu_n\}_{n=1}^{\infty}$ and the spectrum of the string $S_0(I,M)$ generated by $S(I,M)$ coincides with $\{\lambda_n\}_{n=1}^{\infty}$.}

{\bf Proof}  It is enough to repeat arguments in proof of Theorem 1 of \cite{L}, Chapter 7 to show that  
$$
\mathop{\prod}\limits_{n=1}^{\infty}\frac{1-\frac{z}{\lambda_n}}{1-\frac{z}{\mu_n}}:=q(z)
$$
converges for $z\in{\rm Ext} \{\mu_k\}_{k=1}^{\infty}$  and, moreover, $q(z)$ is an $R$-function.   Since $q(z)>0$ for $z\in(-\infty,0)$ the function $q(z)$ is an $S$-function. 
Therefore, $Lq(z)$ is also an $S$-function.

According to M.G. Krein's theorem  (Theorem 11.2 in \cite{KK}, see \cite{DK}, page 252 for the proof) there exists a unique string $S_1(I,M)$   for which $Lq(z)$ is the coefficient of dynamic compliance. Then $\{\mu_k\}_{k=1}^{\infty}$ is the spectrum of this string $S_1(I,M)$ while   $\{\lambda_k\}_{k=1}^{\infty}$   is the spectrum of the string $S_0(I,M)$ with the same $I$ and $M(x)$.  Equation (4.3) implies that the length  of this string  is $Lq(0)=L$.
 Theorem is proved.

In what follows we will say that the string $S_1(I,M)$ existence of which is proved in Theorem 4.2. {\it corresponds to  the data $L$, $\{\mu_k\}_{k=1}^{\infty}$, $\{\lambda_k\}_{k=1}^{\infty}$}.  
\vspace{2mm}

\subsection{Necessary and sufficient conditions for existence of the limit $\lim\limits_{x\to+0}\frac{M(x)}{x^{\alpha}}$ and its calculation via two spectra}

Let limit (\ref{3.12}) exist and be finite and  nonzero. Then (see Subsection 3.6) the second limit in (\ref{3.13}) also exists and (\ref{3.14}) is true, where $B(\alpha)$ is defined by (\ref{3.15}). It follows from (\ref{3.14}) that
\begin{equation}
\label{4.11}
\lim_{x\to+0}x^{-\alpha}M(x)=\left(\frac{1}{B(\alpha)}\lim\limits_{k\to\infty}\mu_k^{-\frac{\alpha}{\alpha+1}}\tau^{(1)}(\mu_k+0)\right)^{-(\alpha+1)}=
\end{equation}
$$
\left(\frac{1}{B(\alpha)}\lim\limits_{k\to\infty}\mu_{k+1}^{-\frac{\alpha}{\alpha+1}}\tau^{(1)}(\mu_{k+1}-0)\right)^{-(\alpha+1)}.
$$
 Therefore, 
\begin{equation}
\label{4.12}
\lim\limits_{k\to\infty}\mu_k^{-\frac{\alpha}{\alpha+1}}\tau^{(1)}(\mu_k+0)=\lim\limits_{k\to\infty}\mu_{k+1}^{-\frac{\alpha}{\alpha+1}}\tau^{(1)}(\mu_{k+1}-0).
\end{equation}
Due to (\ref{4.9})  for each $n\in\mathbb{N}$  we have $\tau^{(1)}(\mu_n+0)=\tau^{(1)}(\mu_{n+1}-0)$. Then (\ref{4.12}) implies that


\begin{equation}
\label{4.13}
\lim\limits_{k\to\infty}\frac{\mu_{k+1}}{\mu_k}=1
\end{equation}
independently of the behavior of $M(x)$ in the exterior of any right neighborhood of $x=0$. 
  
This means that existence of a finite nonzero limit in the left-hand side of (\ref{4.12}) and validity of (\ref{4.13}) are necessary conditions for existence of a finite and nonzero  $\lim\limits_{x\to +0}\frac{M(x)}{x^{\alpha}}$ and of validity of the first of equations (\ref{4.11}). Now we will show that these conditions are sufficient.

First, these conditions imply existence of the limit in the right-hand side of (\ref{4.12}) and validity of equation (\ref{4.12}). Next, since $\tau^{(1)}(\mu_k+0)=\tau^{(1)}(\lambda)=\tau^{(1)}(\mu_{k+1}-0)$,
in each point $\lambda\in(\mu_k,\mu_{k+1})$, the following inequalities are valid:
\begin{equation}
\label{4.14}
\mu_k^{-\frac{\alpha}{\alpha+1}}\tau^{(1)}(\mu_k+0)>\lambda^{-\frac{\alpha}{\alpha+1}}\tau^{(1)}(\lambda)>\mu_{k+1}^{-\frac{\alpha}{\alpha+1}}\tau^{(1)}(\mu_{k+1}-0).
\end{equation}
Since the limits in (\ref{4.12}) are finite and nonzero, inequalities (\ref{4.14}) together with (\ref{4.12}) imply
that 
$$
\lim\limits_{\lambda\to+\infty}\lambda^{-\frac{\alpha}{\alpha+1}}\tau^{(1)}(\lambda)
$$
exists and is finite and nonzero. Then according to Kasahara's theorem  limit  (\ref{3.12}) exists and is finite and nonzero. 
Now we are able to give necessary and sufficient conditions of existence of a nonzero limit (\ref{3.12}) in terms of the spectra:

{\bf Theorem 4.3} {\it Let $M(x)$ be the mass distribution function of a string $S(I,M)$ corresponding to the data $L$, $\{\mu_n\}_{n=1}^{\infty}$ and $\{\lambda_n\}_{n=1}^{\infty}$. Then limit (\ref{3.12}) exists and is finite and nonzero for some $\alpha\in (0,\infty)$ if and only if 

1) the limit
\begin{equation}
\label{4.16}
\lim\limits_{n\to\infty}(\mu_n^{-\frac{\alpha}{\alpha+1}}\tau^{(1)}(\mu_n+0))
\end{equation}
where $\tau^{(1)}(\lambda)$ is given by (\ref{4.9}), exists and is finite and nonzero

2)  (\ref{4.13}) is valid.}

Theorem 4.3 and the first of equations (\ref{4.11}) make possible to find the limit (\ref{3.12}) using the data $L$, $\{\mu_n\}_{n=1}^{\infty}$ and $\{\lambda_n\}_{n=1}^{\infty}$.

{\bf Theorem 4.4}
{\it Let $L>0$ and  two sequences of numbers $\{\mu_n\}_{n=1}^{\infty}$ and $\{\lambda_n\}_{n=1}^{\infty}$ satisfy conditions (\ref{4.10}) and
 behave asymptotically as follows
$$
\lambda_n\mathop{=}\limits_{n\to\infty}\frac{\pi^2 n^2}{b^2}+O(n^{\beta}),  
$$
$$
\mu_n\mathop{=}\limits_{n\to\infty}\frac{\pi^2 (n-1/2)^2}{b^2}+O(n^{\beta}),  
$$
where $\beta\in[0,1)$, $b\in(0,\infty)$ and let $M(x)$ be the mass distribution function of a string $S(I,M)$ corresponding to these data.

Then
\begin{equation}
\label{4.19}
\mathop{\lim}\limits_{x\to +0}M(x)x^{-1}=\frac{1}{L^2\mu_1}\mathop{\prod}\limits_{n=1}^{\infty}\frac{\lambda_n^2}{\mu_n \mu_{n+1}} .
\end{equation}
}

{\bf Proof} It is clear that the product in the right-hand side of (\ref{4.19}) converges and is finite and not zero.  Since the conditions of Theorem 4.2 are satisfied, there exists a  unique string $S(I,M)$ of finite length $L$  with the spectrum of  problem (\ref{3.4}) $\{\mu_n\}_{n=1}^{\infty}$  and  the spectrum of  problem (\ref{3.5}) $\{\lambda_n\}_{n=1}^{\infty}$.  Due to Theorem 4.1 the coefficient of dynamic compliance $T(z)$ of this  string is given by (\ref{4.1}). Thus,

$$
\left(\sqrt{z}T(-z)\right)^{-2}= 
\left(\frac{\mathop{\prod}\limits_{n=1}^{\infty}(1+\frac{z}{\mu_n})}{L\sqrt{z}\mathop{\prod}\limits_{n=1}^{\infty}(1+\frac{z}{\lambda_n})}\right)^2.
$$

Let us evaluate
\begin{equation}
\label{4.20}
\left(\frac{\mathop{\prod}\limits_{n=1}^{\infty}(1+\frac{z}{\mu_n})}{L\sqrt{z}\mathop{\prod}\limits_{n=1}^{\infty}(1+\frac{z}{\lambda_n})}\right)^2=
\frac{1}{L^2}\frac{\mu_1^{(0)}}{\mu_1}\prod\limits_{n=1}^{\infty}\frac{\lambda_n^2}{\mu_n\mu_{n+1}}\prod\limits_{n=1}^{\infty}\frac{\mu_n^{(0)}\mu_{n+1}^{(0)}}{(\lambda_n^{(0)})^2}
\end{equation}
$$
\left(\prod\limits_{n=1}^{\infty}\frac{\mu_n^{(0)}+z}{\mu_n^{(0)}}\right)^2\left(\sqrt{z}\prod\limits_{n=1}^{\infty}\frac{\lambda_n^{(0)}+z}{\lambda_n^{(0)}}\right)^{-2}\left(\prod_{n=1}^{\infty}\left(1+\frac{\mu_n-\mu_n^{(0)}}{\mu_n^{(0)}+z}\right)\right)^2
\left(\prod_{n=1}^{\infty}\left(1+\frac{\lambda_n-\lambda_n^{(0)}}{\lambda_n^{(0)}+z}\right)\right)^{-2}
$$
where $\mu_n^{(0)}=\frac{\pi^2 (n-1/2)^2}{b^2}$ and
$\lambda_n^{(0)}=\frac{\pi^2 n^2}{b^2}$.

Since
$$
\sqrt{z}\prod\limits_{n=1}^{\infty}\frac{\lambda_n^{(0)}+z}{\lambda_n^{(0)}}=\frac{i\sin\sqrt{-z}b}{b}, \ \ \ \
\prod\limits_{n=1}^{\infty}\frac{\mu_n^{(0)}+z}{\mu_n^{(0)}}=\cos\sqrt{-z}b,
$$
we obtain
$$
\left(\frac{\mathop{\prod}\limits_{n=1}^{\infty}(1+\frac{z}{\mu_n})}{L\sqrt{z}\mathop{\prod}\limits_{n=1}^{\infty}(1+\frac{z}{\lambda_n})}\right)^2=
$$
$$
-\frac{\mu_1^{(0)}}{L^2\mu_1}\prod\limits_{n=1}^{\infty}\frac{\lambda_n^2}{\mu_n\mu_{n+1}}\prod\limits_{n=1}^{\infty}\frac{\mu_n^{(0)}\mu_{n+1}^{(0)}}{(\lambda_n^{(0)})^2}(-ib \ cotan \sqrt{-z}b)^2\left(\prod_{n=1}^{\infty}\left(1+\frac{\mu_n-\mu_n^{(0)}}{\mu_n^{(0)}+z}\right)\right)^2
\left(\prod_{n=1}^{\infty}\left(1+\frac{\lambda_n-\lambda_n^{(0)}}{\lambda_n^{(0)}+z}\right)\right)^{-2}
$$
It is clear that 
$$
\lim\limits_{z\to +\infty} ( -i \ {\rm cotan}\sqrt{-z}b)^2=1,
$$
Since the series 
$$
\sum_{n=1}^{\infty}\frac{\lambda_n-\lambda_n^{(0)}}{\lambda_n^{(0)}+z}, \ \ \  \sum_{n=1}^{\infty}\frac{\mu_n-\mu_n^{(0)}}{\mu_n^{(0)}+z}
$$
converge absolutely and uniformly in the neighborhood of $z=+\infty$, we obtain
$$
\lim\limits_{z\to +\infty}\prod_{n=1}^{\infty}\left(1+\frac{\mu_n-\mu_n^{(0)}}{\mu_n^{(0)}+z}\right)=\lim\limits_{z\to +\infty}\prod_{n=1}^{\infty}\left(1+\frac{\lambda_n-\lambda_n^{(0)}}{\lambda_n^{(0)}+z}\right)=1.
$$
We have $\mu_1^{(0)}=\frac{\pi^2}{4b^2}$ and by Wallis formula
$$
\prod\limits_{n=1}^{\infty}\frac{(\lambda_n^{(0)})^2}{\mu_n^{(0)}\mu_{n+1}^{(0)}}=\frac{\pi^2}{4}. 
$$
Finally  we arrive at
$$
\lim\limits_{z\to +\infty}\left(\frac{\mathop{\prod}\limits_{n=1}^{\infty}(1+\frac{z}{\mu_n})}{L\sqrt{z}\mathop{\prod}\limits_{n=1}^{\infty}(1+\frac{z}{\lambda_n})}\right)^2=\frac{1}{L^2\mu_1}\mathop{\prod}\limits_{n=1}^{\infty}\frac{\lambda_n^2}{\mu_n \mu_{n+1}}. 
$$
Now by Kasahara's theorem for $\alpha=1$ with account of Theorem 4.1 we arrive at (\ref{4.19}). Theorem is proved.

\section{Strings with fast growth of spectra}

\setcounter{equation}{0}

In Theorem 4.4 like in \cite{B}, \cite{Sh}, \cite{Sh1} the sequences $\{\lambda_n\}_{n=1}^{\infty}$ and $\{\mu_n\}_{n=1}^{\infty}$ grow roughly speaking as $n^2$. Now let us consider validity of 'Barcilon formula' (\ref{1.4}) in case of more rapid growth of  $\{\lambda_n\}_{n=1}^{\infty}$ and $\{\mu_n\}_{n=1}^{\infty}$.

First we recall a remarkable  M.G. Krein's theorem stated in \cite{Kr2}:

{\bf Theorem 5.1} {\it If a string $S_1(I,M)$ is regular and  $\{\mu_n\}_{n=1}^{\infty}$ is its spectrum then}
$$
\lim\limits_{n\to\infty}\frac{n}{\sqrt{\mu_n}}=\frac{1}{\pi}\int_0^L\sqrt{M^{\prime}(x)}dx.
$$ 
A proof of this theorem can be found in \cite{GK} (Section VI, Theorem 8.1).

This theorem we use to prove the following result.

{\bf Theorem 5.2} {\it If a string  $S_1(I,M)$ of length $L<\infty$ has the spectrum $\{\mu_n\}_{n=1}^{\infty}$  which satisfies
\begin{equation}
\label{5.1}
\lim\limits_{\overline{n\to\infty}}\frac{n}{\sqrt{\mu_n}}=0
\end{equation} 
then $M^{\prime}(x) =0$ almost everywhere on $I$.}

{\bf Proof}
If the string $S_1(I,M)$ is regular the statement of Theorem 5.2 immediately follows from Theorem 5.1.

Now let our string $S_1(I,M)$ be singular. Assume that in contrary to the statement of our theorem the Lebesgue measure of the set $W:=\{x\in I: M^{\prime}(x)>0\}$ is positive. Then there exists $l\in (0,L)$  such that the Lebesgue measure of the set $\hat{W}:=[0,l]\cup W$ is also positive. 

Let us consider the string $S_1(\hat{I},\hat{M})$ where $\hat{I}=[0,l]$ and $\hat{M}(x)=M(x)$  $\forall x\in\hat{I}$. In other words 
$S_1(\hat{I},\hat{M})$ is the part of the string $S_1(I,M)$ located on $[0,l]$. Let $\{\hat{\mu}_n\}_{n=1}^{\infty}$  ($0<\hat{\mu}_1<\hat{\mu}_2<...$)    denote the spectrum of $S_1(\hat{I},\hat{M})$, i.e. the set of points of growth of the main spectral function $\hat{\tau}^{(1)}(\lambda)$ of this string.  The string $S_1(\hat{I},\hat{M})$ is regular, therefore, due to Theorem  5.1 
\begin{equation}
\label{5.2}
\lim\limits_{n\to\infty}\frac{n}{\sqrt{\hat{\mu}_n}}=\frac{1}{\pi}\int_0^l \sqrt{M^{\prime}(x)}dx=\frac{1}{\pi}\int_{\hat{W}} \sqrt{M^{\prime}(x)}dx>0.
\end{equation}
From the definition of a spectral function it  follows that any of  spectral functions of the string $S_1(I,M)$ is a spectral function of the truncated string $S_1(\hat{I},\hat{M})$. In particular,  the main spectral function $\tau^{(1)}(\lambda)$ of $S_1(I,M)$ is a spectral function of $S_1(\hat{I},\hat{M})$. It is enough to compare (\ref{5.1}) with (\ref{5.2}) to clarify  that $\hat{\tau}^{(1)}(\lambda)$ do not coincide with $\tau^{(1)}(\lambda)$. 

According to Theorem 5 of \cite{Kr3} (see also (\cite{Kr4}))
\begin{equation}
\label{5.3}
\tau^{(1)}(\hat{\mu}_{n-1}+0)<\hat{\tau}^{(1)}(\hat{\mu}_{n-1}+0) \ \ n=2,3,...
\end{equation}
and according to Theorem 4 of \cite{Kr3}
\begin{equation}
\label{5.4}
\tau^{(1)}(\hat{\mu}_{n}-0)>\hat{\tau}^{(1)}(\hat{\mu}_{n}-0) \ \ n=1,2,...
\end{equation}
Since $\tau^{(1)}(\hat{\mu}_{n}-0)=\hat{\tau}^{(1)}(\hat{\mu}_{n-1}+0) \ \ n=2,3,...$, we conclude using (\ref{5.3}) and ({\ref{5.4}) that 
$$
\tau^{(1)}(\hat{\mu}_n-0)>\tau^{(1)}(\hat{\mu}_{n-1}+0) \ \ n=2,3,...
$$ 
This means that  there exists at least one point of growth of function $\tau^{(1)}(\lambda)$, i.e. at least one of points $\{\mu_j\}_{j=2}^{\infty}$  on each interval $(\hat{\mu}_{n-1},\hat{\mu}_n)$.

For $n=1$  (\ref{5.4}) gives $\tau^{(1)}(\hat{\mu}_1-0)>\hat{\tau}^{(1)}(\hat{\mu}_1-0)=\hat{\tau}^{(1)}(0)=0=\tau^{(1)}(0)$, i.e. there is at least one point of growth of $\tau^{(1)}(\lambda)$ on the interval $(0,\hat{\mu_1})$. Thus, for each $n\in\mathbb{N}$ the interval $(0, \hat{\mu}_n)$ contains at least $n$ of points of $\{\mu_n\}_{n=1}^{\infty}$. 
Consequently, $\mu_n<\hat{\mu}_n $  $\forall n\in\mathbb{N}$ and according to (\ref{5.2}) 
$$
\lim\limits_{\overline{n\to\infty}}\frac{n}{\sqrt{\mu_n}}>0,
$$
what contradicts (\ref{5.1}). Thus, our assumption that $W$ has positive Lebesgue measure is false. Theorem is proved. 

{\bf Remark} {\it A similar result is true for the spectrum $\{\lambda_k\}_{k=1}^{\infty}$ of a  string $S_0(I,M)$. It follows from the strict interlacing of the spectra  $\{\lambda_k\}_{k=1}^{\infty}$ and  $\{\mu_k\}_{k=1}^{\infty}$ of the strings $S_1(I,M)$ and $S_0(I,M)$, respectively, with the same $I$ and $M$}.

When a nonconstant nondecreasing on $I$ function $M(x)$  is such that $M^{\prime}(x)=0$ almost everywhere the condition of continuity of $M^{\prime}(x)$ used in \cite{B} and \cite{Sh} to prove (\ref{4.19}) is not satisfied. However, Theorem 5.2 together with the following theorem show that there exists a wide class of strings for which (\ref{4.19}) is true despite $M^{\prime}(x)$ is not continuous.

{\bf Theorem 5.3}   
{\it Let $L>0$ and  two sequences of positive numbers $\{\mu_n\}_{n=1}^{\infty}$ and $\{\lambda_n\}_{n=1}^{\infty}$ satisfy conditions (\ref{4.10}) and
$$
\lambda_n=\frac{\pi^4 n^4}{b^4}+O(n^{\beta}),  
$$
$$
\mu_n=\frac{\pi^4(n-1/2)^4}{b^4}+O(n^{\beta}),  
$$
where $b\in (0,\infty)$, $\beta\in [0,3)$.

Then (\ref{4.19}) is true.}

{\bf Proof} 
In our case (\ref{4.20}) remains true with $\mu_n^{(0)}=\frac{\pi^4(n-1/2)^4}{b^4}$ and $\lambda_n^{(0)}=\frac{\pi^4n^4}{b^4}$.
Let us evaluate substituting $\tau=\sqrt{-z}$
$$
\sqrt{z}\prod_{n=1}^{\infty}\frac{\lambda_n^{(0)}+z}{\lambda_n^{(0)}}=i\sqrt{\tau}\prod\limits_{n=1}^{\infty}\frac{\pi^2n^2/b^2-\tau}{\pi^2n^2/b^2}\sqrt{\tau}\prod\limits_{n=1}^{\infty}\frac{\pi^2n^2/b^2+\tau}{\pi^2n^2/b^2}=-\frac{\sin\sqrt{\tau}b\sin\sqrt{-\tau}b}{b^2},   
$$
$$
\prod_{n=1}^{\infty}\frac{\mu_n^{(0)}+z}{\mu_n^{(0)}}=\prod\limits_{n=1}^{\infty}\frac{\pi^2(n-1/2)^2/b^2-\tau}{\pi^2(n-1/2)^2/b^2}\prod\limits_{n=1}^{\infty}\frac{\pi^2(n-1/2)^2/b^2+\tau}{(n-1/2)^2}=\cos\sqrt{\tau}b\cos\sqrt{-\tau}b.
$$
Using these formulae we obtain from (\ref{4.20})
$$
\left(\frac{\mathop{\prod}\limits_{n=1}^{\infty}(1+\frac{z}{\mu_n})}{L\sqrt{z}\mathop{\prod}\limits_{n=1}^{\infty}(1+\frac{z}{\lambda_n})}\right)^2=
$$
$$
-\frac{\mu_1^{(0)}}{L^2\mu_1}\prod\limits_{n=1}^{\infty}\frac{\lambda_n^2}{\mu_n\mu_{n+1}}\prod\limits_{n=1}^{\infty}\frac{\mu_n^{(0)}\mu_{n+1}^{(0)}}{(\lambda_n^{(0)})^2}(b^2 \ cotan \sqrt{\tau}b\ cotan\sqrt{-\tau}b)^2
$$
$$
\left(\prod_{n=1}^{\infty}\left(1-\frac{\mu_n-\mu_n^{(0)}}{\mu_n^{(0)}+z}\right)\right)^2
\left(\prod_{n=1}^{\infty}\left(1-\frac{\lambda_n-\lambda_n^{(0)}}{\lambda_n^{(0)}+z}\right)\right)^{-2}
$$

It is clear that 
$$
\lim\limits_{\tau\to \pm i\infty} ( {\rm cotan}\sqrt{\tau}b \ {\rm cotan}\sqrt{-\tau}b)^2=1.
$$
Since the series 
$$
\sum_{n=1}^{\infty}\frac{\lambda_n-\lambda_n^{(0)}}{\lambda_n^{(0)}+z}, \ \ \  \sum_{n=1}^{\infty}\frac{\mu_n-\mu_n^{(0)}}{\mu_n^{(0)}+z}
$$
converge absolutely and uniformly in the neighborhood of $z=+\infty$
$$
\lim\limits_{z\to +\infty}\prod_{n=1}^{\infty}\left(1-\frac{\mu_n-\mu_n^{(0)}}{\mu_n^{(0)}-\lambda}\right)=\lim\limits_{z\to+\infty}\prod_{n=1}^{\infty}\left(1-\frac{\lambda_n-\lambda_n^{(0)}}{\lambda_n^{(0)}-\lambda}\right)=1
$$
and
$$
\prod\limits_{n=1}^{\infty}\frac{(\lambda_n^{(0)})^2}{\mu_n^{(0)}\mu_{n+1}^{(0)}}=\frac{\pi^4}{16}, \ \  \ \ \mu_1^{(0)}=\frac{\pi^4}{16b^4}.
$$

Therefore, we have
$$
\lim\limits_{z\to +\infty}\left(\frac{\mathop{\prod}\limits_{n=1}^{\infty}(1+\frac{z}{\mu_n})}{L\sqrt{z}\mathop{\prod}\limits_{n=1}^{\infty}(1+\frac{z}{\lambda_n})}\right)^2= 
\frac{1}{L^2\mu_1}\mathop{\prod}\limits_{n=1}^{\infty}\frac{\lambda_n^2}{\mu_n \mu_{n+1}}.
$$
Theorem is proved.


\subsection*{Acknowledgments}
The second author was partly supported by the Swiss National Science
Foundation through grant IZ73Z0-128135.

\end{document}